# Voltage induced friction in a contact of a finger and a touchscreen with a thin dielectric coating

Valentin L. Popov and Markus Heß

Technische Universität Berlin, Str. des 17. Juni 135, 10623 Berlin, Germany

**Abstract.** We show that application of voltage between two electrically conducting bodies separated by a thin dielectric layer is equivalent to an adhesive contact with an effective separation energy depending quadratically on the applied stress. Under assumption of Coulomb friction in the contact interface, a closed form equation for the friction force is derived.

## 1. Introduction

Friction is an important quantity influencing the tactile perception of humans and other biological species [1]. It is known that the level of friction can be changed by application of a voltage between the finger and the substrate. This effect was experimentally investigated almost one century ago, in 1917-1923, by Johnsen and Rahbek [2] and became recently again a topic of high current interest due to applications of electrohaptics in robotics and in systems for shape recognition for blind persons [3], [4]. However, the detailed mechanism of the influence of voltage on friction is still not understood [4]. In the present paper we use the initial idea of Johnsen and Rahbek to develop a quantitative model for voltage induced friction. Johnsen and Rahbek write: "In the course of some experiments in telephony, the authors discovered in 1917 that in some cases an appreciable adhesion occurs between two bodies in contact with each other, when a difference of electrical potential exists between them." [2]. In the present paper we will show, that a contact of conducting bodies can indeed be considered as adhesive contact with the work of adhesion depending on the applied voltage. We use this idea and the recent understanding of friction in adhesive contacts to derive a simple explicit equation for the force of friction.

## 2. Problem statement and model description

The background physical system which we would like to describe is a finger in a contact with a flat rigid screen consisting of conducting substrate and rigid dielectric coating. We model the finger as conductive elastic axisymmetrical paraboloid with the shape function

$$f(r) = \frac{r^2}{2R}. \tag{1}$$

The assumption of axisymmetricity is not essential for the following consideration but simplifies the parametrization of the contact and allows more concise presentation of the main idea. The considered model is schematically shown in Fig. 1. The external load can be described either with the indentation depth $d$ (which in adhesive case can be both positive and negative) and the normal force $F_N$. We define the contact radius $a$ as the radius of contact area with the dielectric coating. Further, we introduce the distance $h(r)$ between the surface of the "finger" and the *conducting substrate*. Inside the contact area, this distance is equal to the thickness $h_0$ of the dielectric coating. If a voltage $V$ is applied between the conducting substrate and the finger, the surfaces of these bodies become electrically charged, as in a plate capacitor and attract each other. As the charge in conducting bodies is distributed in the thin surface layer [5], we have to do with attractive interaction directly between two surfaces that is with the classical problem of adhesion. In the present paper, we will assume that solution of the corresponding adhesive contact problem is given in good approximation by the JKR theory, which means that the characteristic range of interaction is much smaller than the indentation depth. The most important parameter of adhesive interaction – the specific work of adhesion can be easily specified in explicit form.

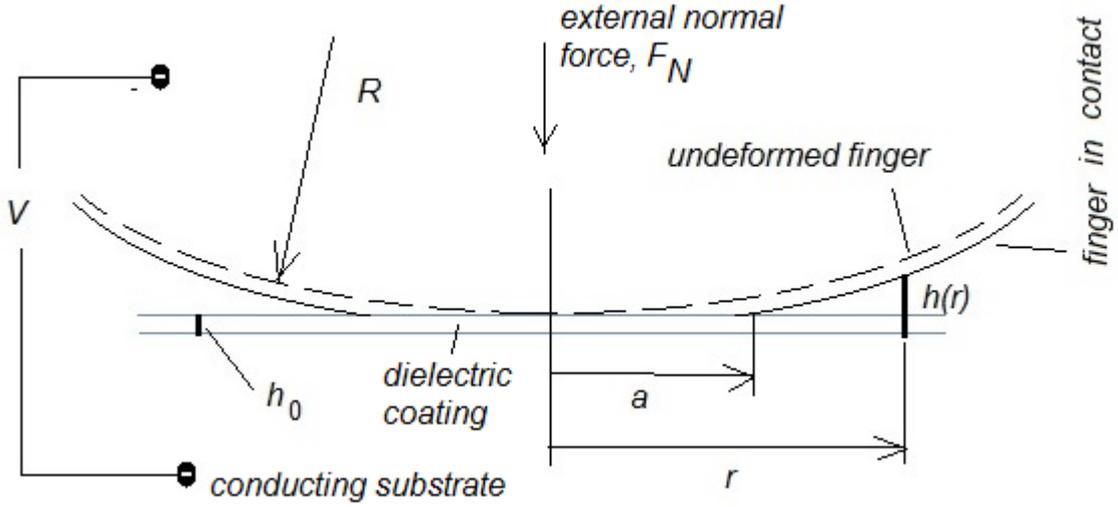

Fig. 1 Schematic presentation of the model. An elastic axis-symmetric paraboloid (radius of curvature $R$) is in contact with the rigid plane counter body consisting of conducting substrate and dielectric hard coating with thickness $h_0$. Between the substrate and the finger, a voltage $V$ is applied.

## 3. Electro adhesion

Consider two elements of the surface of the "finger" and the substrate opposing each other, each having area $\Delta A$. They build a capacitor with the plate spacing $h$, which is connected to a voltage source, which ensures the constancy of the voltage regardless of the (variable) plate spacing. In our case, the space between the conducting surfaces is filled partially with material having relative permittivity $\varepsilon$ (the dielectric layer with thickness $h_0$) while the rest is filled with air, with relative permittivity approximately equal to one. This structure can be considered as a series connection of partial capacities, so that the total capacitance is given by

$$\frac{1}{C} = \frac{1}{C_1} + \frac{1}{C_2}, \qquad (2)$$

with

$$C_1 = \varepsilon_0 \varepsilon \frac{\Delta A}{h_0} \quad \text{and} \quad C_2 = \varepsilon_0 \frac{\Delta A}{h - h_0}, \qquad (3)$$

where $\varepsilon_0$ is the vacuum permittivity. Inserting (3) into (2) returns the total capacitance

$$C = \frac{\varepsilon_0 \varepsilon \Delta A}{h_0 + \varepsilon (h - h_0)}. \qquad (4)$$

The attractive electric force acting between the plates is given by the Eq.

$$\left|F_{el}\right| = \frac{1}{2} C^2 V^2 \left|\frac{\mathrm{d}}{\mathrm{d}h}\left(\frac{1}{C(h)}\right)\right|. \qquad (5)$$

With account of (4), it follows

$$\left|F_{el}(h)\right| = \frac{1}{2} V^2 \frac{\varepsilon_0 \varepsilon^2 \Delta A}{\left(h_0 + \varepsilon (h - h_0)\right)^2}. \qquad (6)$$

The normal component of attractive stress is thus

$$\sigma(h) = \frac{F_{el}(h)}{\Delta A} = \frac{1}{2}V^2 \frac{\varepsilon_0 \varepsilon^2}{\left(h_0 + \varepsilon(h - h_0)\right)^2}. \tag{7}$$

In particular, in direct contact ($h = h_0$), the stress is equal to

$$\sigma_0 = \sigma(h_0) = \frac{1}{2}\varepsilon_0 \varepsilon^2 \frac{V^2}{h_0^2}. \tag{8}$$

For the specific work of adhesion we get

$$\Delta\gamma = \int_{h_0}^{\infty} \sigma(h) \, dh = \frac{1}{A}\int_{h}^{\infty} F_{el}(h) \, dh = \frac{\varepsilon_0 \varepsilon}{2h_0} V^2. \tag{9}$$

The considered contact is of course an adhesive contact with adhesive interaction (7). Assuming large value of the Tabor parameter [6], we can apply to this contact the theory of Johnson, Kendall and Roberts (JKR) [7] to find the relations between the normal force, the indentation depth and the contact radius. The only difference from the classical JKR-theory is explicit dependence of the work of adhesion on the applied voltage, given by Eq. (9). This means, that applying voltage, one can control adhesion.

## 4. Friction under voltage

Friction in an adhesive contact is a problem which was recently controversially discussed. In this Section we use the model considered in [8] and reproduced with improved derivation in [9]. In [8] and [9], it was shown that the force of friction in the sliding state is given by the Equation

$$F_x = \mu\left(\pi a^2 \sigma_0 + F_{N,\text{adh}}(a)\right), \tag{10}$$

where $a$ is the contact radius in the adhesive contact corresponding to the normal force $F_{N,\text{adh}}$, $\sigma_0$ is the adhesive stress in the direct contact (in our case given by the Eq. (8)), and $\mu$ the coefficient of friction in the interface between the finger and the dielectric layer.

In the limiting case of JKR, Eq. (10) can be written explicitly as

$$F_x = \mu\left(\pi a^2 \frac{1}{2}\varepsilon_0 \varepsilon^2 \frac{V^2}{h_0^2} + \frac{4}{3}\frac{E^* a^3}{R} - \sqrt{\frac{8\pi a^3 E^* \varepsilon_0 \varepsilon}{2h_0} V^2}\right), \tag{11}$$

where the contact radius is determined by the JKR-Equation

$$F_{N,\text{adh}}(a) = \frac{4}{3}\frac{E^* a^3}{R} - \sqrt{\frac{8\pi a^3 E^* \varepsilon_0 \varepsilon}{2h_0} V^2}. \tag{12}$$

Here

$$E^* = \frac{E}{1 - \nu^2} \tag{13}$$

where $E$ is the Young modulus of the finger and $\nu$ its Poisson number.

Note that in the absence of the external normal force ($F_{N,\text{adh}} = 0$) there will be still some force of friction, $F_x = \mu \pi a^2 \sigma_0$, where $a$ is the contact radius in adhesive contact without external load. In this case Eq. (12) and (11) lead to the result

$$F_x = \frac{1}{2}\mu\pi^{\frac{5}{3}}\left(\frac{9}{4}\right)^{\frac{2}{3}} R^{\frac{4}{3}} E^{*-\frac{2}{3}} \varepsilon^{\frac{8}{3}} \varepsilon_0^{\frac{5}{3}} h_0^{-8/3} V^{10/3}, \quad \text{for } F_{N,\text{adh}} = 0\,. \qquad (14)$$

## 5. Conclusions

In the present paper we followed the ingenious insight by A. Johnsen and K. Rahbek considering the voltage induced attraction of conducting surfaces as a sort of adhesion. The contact of two conducting bodies separated by a thin dielectric layer is indeed equivalent to the adhesive contact problem with the work of separation depending on the applied voltage. Recent advances in understanding of adhesive contacts with friction allowed for a simple analytical solution for the force of friction, assuming Coulomb friction in the immediate interface between the bodies. The force of friction is given in general case by Eq. (10) and in the JKR limit by Eq. (11).

## 6. References

[1] M.J. Adams, S.A. Johnson, P. Lefèvre, V. Lévesque, V. Hayward, T. André, J.-L. Thonnard, Finger pad friction and its role in grip and touch, *Journal of The Royal Society Interface*, 10 (80), 20120467, 2013

[2] A. Johnsen, K. Rahbek, A physical phenomenon and its applications to telegraphy telephony etc, *Electrical Engineers Journal of the Institution*, 61 (320), 713-725, 1923.

[3] C.D. Shultz, M.A. Peshkin, and J. E. Colgate, Surface haptics via electroadhesion: Expanding electrovibration with Johnsen and Rahbek, *2015 IEEE World Haptics Conference* (WHC), Jun. 2015.

[4] Y. Vardar, B. Güçlü, and C. Basdogan, Effect of Waveform in Haptic Perception of Electrovibration on Touchscreens, *Lecture Notes in Computer Science*, pp. 190–203, 2016.

[5] L.D. Landau & E.M. Lifshitz, Electrodynamics of Continuous Media (Volume 8 of A Course of Theoretical Physics), Pergamon Press, 1960.

[6] D. Tabor, Surface Forces and Surface Interactions, J. Colloid Interface Sci. 58(1), 2-13 1977.

[7] K.L. Johnson, K. Kendall, A.D. Roberts, Surface Energy and the Contact of Elastic Solids. *Proceedings of the Royal Society of London*, Series A, 324, 301–313, 1971.

[8] V.L. Popov, A.V. Dimaki, Friction in an adhesive tangential contact in the Coulomb-Dugdale approximation, *The Journal of Adhesion*, 93 (14), 1131-1145, 2017.

[9] V.L. Popov, M. Heß, and E. Willert, Handbuch der Kontaktmechanik, Springer, 341 p, 2018.